# The "Entropy of Knowledge" (EoN): Complexity, Uncertainty, and the Quest for Scientific Knowledge

Babu George, PhD
Alcorn State University, USA

**Abstract**

This paper explores the concept of "entropy of knowledge" (EoN) as a framework for understanding the challenges and complexities of scientific discovery. Drawing from principles in thermodynamics and information theory, I propose that the pursuit of knowledge is characterized by a natural tendency towards disorder, uncertainty, and false conclusions. My central argument is that this entropy of knowledge is not merely an obstacle to be overcome but a fundamental feature of the scientific process, necessary for the exploration of new ideas and the ultimate attainment of truth. The implications of this perspective for the management of scientific inquiry is considered and the same suggests a cyclical approach that balances periods of openness and disorder with phases of consolidation and consensus-building. An attempt is made to situate the EoM model within the broader context of theories of scientific progress, drawing connections to Thomas Kuhn's concept of paradigm shifts and the notion of punctuated equilibrium in the history of science.

**Introduction**

The quest for understanding is a defining feature of human experience. Throughout history, individuals and societies have sought to make sense of the world around them, to uncover the hidden laws and principles that govern the behavior of nature and the workings of the cosmos (Natal et al, 2021). This pursuit of knowledge has driven the development of science, philosophy, and technology, shaping the course of human civilization and transforming our relationship with the universe (Bolisani et al., 2018). However, the path to knowledge is rarely straight or smooth (Rooney & McKenna, 2007). The world is a complex and often confusing place, filled with countless variables, interactions, and uncertainties. Even the most basic questions about the nature of reality or the origins of life remain hotly debated, despite centuries of investigation and discovery (Baranger, 2000). In many cases, the more we learn, the more we realize how much we don't know, and how much there is still to uncover (Ayers, 1997. This complexity poses a fundamental challenge for the pursuit of knowledge. *How can we hope to make progress in understanding the world when there are so many possible explanations, so many potential dead ends and false leads? How can we separate signal from noise, truth from error, in a landscape of constantly shifting evidence and evolving theories?*

To grapple with these questions, we may turn to entropy, a powerful idea that has emerged from the fields of thermodynamics and information theory (Chumachenko et al., 2022). In its most basic sense, entropy is a measure of disorder or uncertainty in a system (Crofts, 2007). In thermodynamics, it refers to the amount of energy that is unavailable for useful work, the degree of randomness and chaos in a physical process. In information theory, it describes the level of unpredictability or randomness in a message or signal. Despite their different contexts, these two conceptions of entropy share a common theme: the tendency of systems to move from states of order and predictability to states of disorder and uncertainty.

In the physical world, this is exemplified by the second law of thermodynamics, which states that the total entropy of an isolated system always increases over time. In the realm of information, it is captured by Shannon's theory of communication, which shows how the entropy of a message sets fundamental limits on our ability to compress or transmit it.

Entropy, with its emphasis on disorder and uncertainty, provides a powerful lens for understanding the challenges of knowledge and scientific discovery (Bawden & Robinson, 2018). Just as physical systems tend towards states of greater entropy, we can see the pursuit of knowledge as a battle against the natural drift towards confusion and misinformation (Fuller, 2019). Every false conclusion, every discarded theory, every null result can be seen as an increase in the entropy of our understanding, a step away from the ordered state of perfect knowledge. But entropy is not just a metaphor for the difficulties of scientific progress. It can also offer insights into the very nature of the discovery process. In the same way that the second law of thermodynamics suggests that some level of inefficiency and waste is inevitable in any real-world process, the entropy of knowledge (EoN) suggests that some degree of error and confusion is unavoidable in the pursuit of truth. Moreover, just as entropy has led to profound insights in fields ranging from statistical mechanics to computer science, applying it to the domain of knowledge could yield new understandings of the scientific process and the nature of intellectual progress. By bringing the EoN, we may find new ways to navigate the complexities of discovery and to make sense of the twists and turns on the path to understanding.

*In this backdrop, I will try to bring home the following propositional ideas:*

1. The pursuit of knowledge is inherently characterized by a tendency towards disorder, uncertainty, and false conclusions, which I term the EoN.

2. The EoN is a fundamental feature of the scientific process, reflecting the complexity of the world and the limitations of human understanding, rather than a mere obstacle to be overcome.

3. The history of science demonstrates the pervasive presence of entropy, as evidenced by the prevalence of false starts, dead ends, and paradigm shifts in the evolution of scientific understanding.

4. Effective navigation of the EoN requires active management and intervention, including the cultivation of openness and diversity, the practice of rigorous self-correction, and the strategic allocation of resources.

5. The concept of EoN is consistent with and complementary to existing theories of scientific progress, such as Thomas Kuhn's notion of paradigm shifts and the theory of punctuated equilibrium in evolutionary biology.

Integrating the EoN entails a shift in the way we approach scientific inquiry, emphasizing the value of uncertainty, the iterative nature of discovery, and the importance of intellectual humility and adaptability. The insights and strategies suggested by the EoN framework have potential applications beyond the realm of science, including in education, policy-making, and public discourse.

**The Entropy of Knowledge Explained**

The entropy of knowledge is a novel perspective that draws from the principles of thermodynamics and information theory to characterize the inherent challenges and complexities of scientific discovery. Knowledge may be viewed as a system that naturally tends towards disorder, uncertainty, and false conclusions (Boisot & Canals, 2004). This is much like the entropy of a physical system.

*We may define EoN as a measure of the uncertainty, confusion, or misinformation present in a given body of knowledge or field of inquiry.* High entropy corresponds to a state of knowledge that is highly disordered, with many competing theories, conflicting evidence, and dead-end research paths. Low entropy, on the other hand, characterizes a body of knowledge that is Ill-ordered, with a clear consensus on key principles, strong empirical support, and a coherent theoretical framework.

The history of science is replete with examples of the natural drift towards disorder and uncertainty in the pursuit of knowledge (Best, 1991; Jakimowicz, 2020). From the geocentric model of the universe to the theory of phlogiston, scientific understanding has often been clouded by false assumptions, misleading data, and flawed interpretations (Chen & Song, 2017). Even in the most rigorous and Ill-established fields, there are always open questions, unresolved anomalies, and areas of active debate. This tendency towards entropy can be understood as a consequence of the inherent complexity of the world and the limitations of human cognition (Brier, 2010). The universe is a vast and intricate system, with countless variables and interactions that defy simple explanation. Our attempts to make sense of this complexity are always partial and provisional, based on incomplete information and shaped by our own biases and assumptions. The process of scientific inquiry itself is subject to various sources of noise and uncertainty (Modis, 2022). Experimental measurements are always subject to error and variability, while theoretical models are necessarily simplifications of reality. The social and institutional context of science, with its competing incentives and human failings, can also introduce distortions and biases into the research process.

False conclusions and wrong turns are not merely obstacles to scientific progress, but necessary and inevitable steps on the path to truth (Georgescu-Roegen, 1971; Gilder, 2013). Just as the second law of thermodynamics implies that some level of inefficiency and waste is unavoidable in any real-world process, the EoN suggests that some degree of error and confusion is essential to the process of discovery. This idea can be seen in the iterative nature of the scientific method, which relies on a cycle of hypothesis, experimentation, and revision to gradually refine our understanding of the world (Jacobs, 2023). Each false hypothesis or failed experiment provides valuable information, ruling out certain possibilities and narrowing the search space for the correct explanation. Without these wrong turns and dead ends, science would be a much less efficient and effective process. The necessity of entropy in knowledge can be seen as a reflection of the fundamental creativity and open-endedness of the scientific enterprise. If the path to truth were a straightforward, linear process, with no room for surprise or serendipity, science would be a much less dynamic and innovative endeavor (Bratianu, C., & Bejinaru, 2019; Sachs, 2008). It is precisely the unpredictability and uncertainty of the research process that allows for the emergence of novel ideas and unexpected breakthroughs.

The history of science is rich with examples that (unintentionally) illustrate EoN. One classic case is the development of our understanding of the nature of light. For centuries, scientists grappled with competing theories about whether light was a particle or a wave, with each camp marshaling evidence and arguments in support of their view. It wasn't until the early 20th century, with the advent of quantum mechanics, that a more complete and coherent understanding emerged, one that reconciled the particle and wave aspects of light as complementary features of a single underlying reality. Another example can be found in the field of medicine, where the development of effective treatments has often been hampered by false assumptions and flawed theories. From the ancient Greek notion of the four humors to the more recent idea of "bad blood," medical science has been shaped by a succession of incorrect models and misguided practices. It is only through the gradual accumulation of empirical evidence and the constant revision of theoretical frameworks that we have arrived at our current understanding of health and disease (Maisseu & Voss, 1995).

These examples illustrate the central role of entropy in the process of scientific discovery. They show how progress in science is not a smooth, linear accumulation of facts, but a messy, iterative process of trial and error, of false starts and dead ends. They demonstrate how the very complexity and uncertainty that makes science challenging is also what makes it so powerful and transformative.

**Managing the EoN**

While the EoN is a natural and necessary feature of the scientific process, it also poses certain risks and challenges that must be carefully managed. If left unchecked, the proliferation of false conclusions, conflicting theories, and dead-end research paths can lead to a state of intellectual paralysis, where the signal of true discovery is drowned out by the noise of misinformation and confusion.

One potential risk is the erosion of public trust in science. When the scientific community appears to be in a constant state of disagreement and contradiction, with no clear consensus on even basic questions, it can undermine the credibility and authority of scientific institutions in the eyes of the public. This can create a fertile ground for the spread of pseudoscience, conspiracy theories, and anti-intellectual attitudes. Another risk is the waste of limited resources on unproductive research directions. In a world of finite funding and human capital, the pursuit of false leads and dead ends can divert resources away from more promising avenues of inquiry. This can slow the overall pace of scientific progress and delay the development of important technologies and innovations.

Given these potential risks, there is a strong case to be made for active management and intervention in the EoN. While some level of disorder and uncertainty is inevitable and even desirable, too much can be counterproductive and damaging. The goal of knowledge entropy management is not to eliminate uncertainty altogether, but rather to find a balance that optimizes the efficiency and effectiveness of the scientific enterprise. This active management can take many forms. It can involve the development of more rigorous standards of evidence and methodological guidelines to help distinguish reliable findings from spurious ones. It can include the creation of institutional mechanisms for the early detection and correction of errors, such as peer review, replication studies, and post-publication critique. It can also involve the cultivation of a scientific culture that values openness, transparency, and self-correction over dogmatism and defensiveness.

One approach to managing the EoN is a cyclical model that alternates between periods of openness and consolidation. During the open phase, a wide range of ideas and hypotheses are allowed to proliferate, with minimal constraints or filtering. This allows for the maximum exploration of the possibility space and the potential for serendipitous discoveries. However, this open phase is followed by a period of consolidation, where the most promising ideas are selected and refined through a process of rigorous testing and critique. This consolidation phase serves to reduce the EoN, creating a more ordered and coherent body of understanding that can serve as a foundation for further inquiry. This cyclical approach can be applied at multiple scales, from individual research projects to entire scientific fields. It allows for the benefits of creative exploration while still imposing the necessary discipline and quality control. It recognizes the importance of both diversity and selection in the evolution of scientific knowledge.

The effective management of knowledge entropy also requires the support of robust institutions, collaborative networks, and shared standards of evidence. Scientific institutions, such as universities, research centers, and funding agencies, play a critical role in setting the incentives and norms that shape the conduct of research. They can promote practices that minimize bias and error, such as data sharing, pre-registration of studies, and the use of blind analysis methods. Collaborative networks, both within and across disciplines, are also essential for the management of knowledge entropy. By bringing together diverse perspectives and expertise, these networks can help to identify and correct errors, to synthesize disparate findings into coherent theories, and to disseminate reliable knowledge to a wider audience.

The development and adoption of clear standards of evidence is crucial for the effective management of knowledge entropy. These standards should be based on the best available methodologies and practices and should be continuously updated considering new insights and capabilities. They should be applied consistently and transparently across different research domains, while still allowing for flexibility and adaptation to specific contexts.

By attending to these institutional, collaborative, and evidential dimensions, the scientific community can create a more robust and resilient ecosystem for the pursuit of knowledge. It can employ the creative power of entropy while still maintaining the necessary checks and balances to ensure the reliability and integrity of the scientific enterprise. It is to be reiterated that the goal is not to eliminate the EoN, but to navigate it in a way that maximizes the efficiency and effectiveness of the discovery process.

**The EoN and Theories of Scientific Progress**

The EoN resonates with several influential theories about the nature of scientific progress. One of the most prominent of these is Thomas Kuhn's idea of paradigm shifts, introduced in his seminal work "The Structure of Scientific Revolutions." According to Kuhn, the history of science is not a linear accumulation of facts and theories, but rather a series of alternating phases of "normal science" and "revolutionary science." During periods of normal science, researchers work within an established paradigm, a shared framework of assumptions, methods, and exemplars that guide their inquiries. This paradigm allows for efficient puzzle-solving and incremental progress, but it also constrains the kinds of questions that can be asked and the types of answers that can be accepted.

However, as anomalies and inconsistencies accumulate within a paradigm, the entropy of knowledge increases. The existing framework becomes less and less able to account for new observations and insights, leading to a state of crisis. It is at this point that a revolutionary phase of science can emerge, marked by the development of radically new theories and the overthrow of the old paradigm.

This pattern of alternating stability and upheaval in the history of science can be seen as a form of punctuated equilibrium, a concept borrowed from evolutionary biology. Just as species can undergo long periods of relative stasis, punctuated by brief bursts of rapid change, scientific knowledge can experience extended phases of incremental refinement, interrupted by sudden leaps and transformations. The EoN plays a key role in these dynamics of scientific change. During periods of normal science, entropy is relatively low, as the dominant paradigm provides a coherent and ordered framework for research. As we have seen before, as anomalies and contradictions accumulate, the entropy increases, creating the conditions for a revolutionary shift. This perspective suggests that the growth of knowledge is not a smooth, continuous process, but rather a complex, non-linear one characterized by thresholds, phase transitions, and emergent phenomena. It highlights the importance of crisis and conflict in the evolution of scientific understanding, and the necessity of periodic upheavals in the structure of knowledge.

One application of this is in the development of formal models of scientific change that incorporate the EoN as a key variable. These models could help to elucidate the conditions and mechanisms that drive the transitions between normal and revolutionary science, and to predict the likelihood and timing of future paradigm shifts. Another opportunity is in the empirical study of the history and sociology of science, using the EoN as a lens for analyzing the development of specific fields or research programs. By measuring the entropy of a body of knowledge at different points in time, we could gain insights into the dynamics of scientific progress, the emergence of new theories and methods, and the factors that contribute to the success or failure of different research trajectories. Integrating EoN into theories of scientific progress can enrich our understanding of the complex, non-linear nature of the discovery process. It can help us to appreciate the value of disorder and uncertainty as drivers of innovation and change, while also highlighting the need for active management and intervention to ensure the efficiency and reliability of the scientific enterprise.

**Interdisciplinary Collaborations**

The EoN framework highlights the importance of interdisciplinary collaboration in navigating the complexity of the knowledge landscape. By bringing together perspectives and approaches from different fields, researchers can gain new insights, challenge established assumptions, and develop more robust and innovative solutions to complex problems (Ribeiro et al., 2021; Schmitt, 2019). For example, the application of complexity science and network theory to the study of knowledge production could provide valuable tools for modeling and analyzing the dynamics of scientific communities, the emergence of new ideas and fields, and the spread of information and influence within and across disciplines. Similarly, insights from the social sciences, such as science and technology studies, the sociology of knowledge, and the philosophy of science, can help to situate the EoN framework within a broader understanding of the social, cultural, and political contexts that shape the production and circulation of knowledge. Interdisciplinary collaboration can also help to address some of the limitations and challenges of the EoN framework itself. By engaging with researchers from different fields, proponents of this framework can refine and expand their concepts and methods, incorporate new forms of evidence and analysis, and situate their work within a more comprehensive understanding of the nature of knowledge and the challenges of scientific progress.

To facilitate interdisciplinary collaboration, research institutions and funding bodies can create incentives and opportunities for researchers to work across disciplinary boundaries, such as through joint funding schemes, interdisciplinary research centers, and cross-disciplinary training programs. By fostering a culture of openness, curiosity, and intellectual risk-taking, these initiatives can help to unlock the creative potential of the research community and accelerate the generation of new ideas and approaches.

**Implications and Future Directions**

The EoN has significant implications for how we approach scientific inquiry and the pursuit of understanding. It suggests that uncertainty and complexity are not merely obstacles to be overcome, but fundamental features of the knowledge landscape that must be actively engaged and navigated. This perspective calls for a shift in the way we think about the goals and methods of science. Rather than seeking absolute certainty or definitive answers, we should welcome the inherent provisionality and incompleteness of our understanding. We should recognize that the path to knowledge is always winding and iterative, marked by false starts, dead ends, and unexpected detours. At the same time, we should not mistake uncertainty for futility or relativism. The EoN framework suggests that the scientific process, while messy and non-linear, is ultimately a self-correcting and progressive one. By constantly testing and revising our theories, by seeking out anomalies and engaging with critique, we can gradually reduce the entropy of our understanding and arrive at more reliable and comprehensive models of reality.

For individual researchers, the EoN framework suggests the importance of cultivating intellectual humility, openness to alternative perspectives, and a willingness to question one's own assumptions. Researchers should be encouraged to engage actively with the uncertainty and complexity of their fields, to seek out feedback and criticism, and to view their work as part of an ongoing, collective process of knowledge production. Research institutions and universities can support this by fostering a culture of openness, diversity, and interdisciplinary collaboration. This may involve creating spaces and platforms for researchers from different fields to interact and exchange ideas, as well as promoting practices such as data sharing, open access publishing, and public engagement with science. Funding bodies can play a key role in navigating the EoN by supporting research that is innovative, risk-taking, and responsive to emerging challenges and opportunities. This may require moving beyond narrow, discipline-specific funding models and towards more flexible, problem-oriented approaches that encourage collaboration and experimentation. Funding agencies can also prioritize the development of robust research infrastructures, such as data repositories, collaborative platforms, and tools for monitoring and evaluating research quality and impact. By investing in these collective resources and capabilities, funders can help to create a more resilient and adaptive research ecosystem that is better equipped to navigate the complexity of the knowledge landscape.

To effectively navigate the complexity of the knowledge landscape, it is essential to foster a scientific culture that values openness, diversity, and active management (Arnheim, 1971; Daly & Farley, 2011; Parrinder, 2015). This means creating an environment that encourages the free exchange of ideas, the frank admission of uncertainty and error, and the continuous updating of assumptions and beliefs considering new evidence (Marks, 2009). It also means actively promoting diversity and inclusivity in the scientific community, recognizing that a multiplicity of perspectives and approaches is essential for the health and vitality of the research enterprise. By bringing together individuals from different backgrounds, disciplines, and ways of thinking, we can enhance the creative potential of science and accelerate the process of discovery. The EoN framework highlights the importance of active management and intervention in the research process. This includes the development of robust methods for the detection and correction of errors, the promotion of transparency and reproducibility in research practices, and the strategic allocation of resources to the most promising and impactful areas of inquiry.

While EoN has been developed primarily in the context of scientific research, it has potential applications and implications that extend far beyond the realm of science. The insights and strategies suggested by this framework could be valuable in any domain where the pursuit of knowledge and understanding is central, such as education, policy-making, or public discourse. In education, for example, embracing the EoN could lead to pedagogical approaches that emphasize the value of uncertainty, the iterative nature of learning, and the importance of critical thinking and intellectual humility. Students could be encouraged to engage actively with the complexity of knowledge, to question their assumptions, and to see the process of inquiry as an ongoing and collaborative endeavor. In the policy realm, the EoN perspective could inform the design of decision-making processes that are more robust to uncertainty and responsive to new evidence. It could highlight the importance of diverse perspectives, interdisciplinary collaboration, and the continuous updating of policies in light of changing circumstances and understanding. In public discourse, EoN could help to promote a more nuanced and constructive approach to complex and controversial issues.

**Conclusion**

The concept of EoN, as developed in this paper, offers a novel and compelling framework for understanding the challenges and complexities of scientific discovery. Drawing from the principles of thermodynamics and information theory, we have argued that the pursuit of knowledge is characterized by a natural tendency towards disorder, uncertainty, and false conclusions. This EoN is not merely an obstacle to be overcome, but a fundamental feature of the scientific process that reflects the inherent complexity of the world and the limitations of human understanding. We have seen how the history of science is replete with examples of false starts, dead ends, and paradigm shifts, illustrating the non-linear and iterative nature of the discovery process. At the same time, we emphasized the importance of active management and intervention in navigating the EoN. By promoting a culture of openness, diversity, and rigorous self-correction, and by developing robust methods for the detection and correction of errors, we can harness the creative potential of entropy while mitigating its risks and pitfalls.

A summary depiction of the dynamics of the entropic nature of knowledge discovery is given below in figure 1.

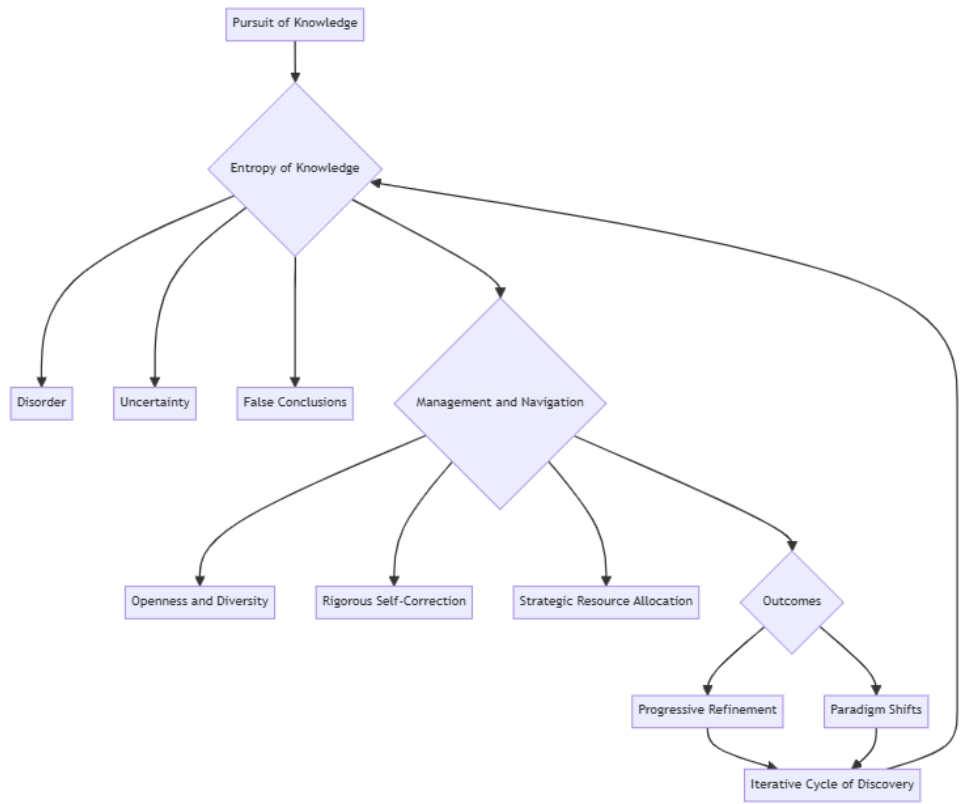

Figure 1: The entropic processes involved in knowledge discovery

The EoN framework highlights the ongoing challenge and necessity of navigating the complexities of the discovery process. As our understanding of the world continues to grow and evolve, so too will the landscape of knowledge become ever more intricate and multifaceted. New questions will emerge, new anomalies will arise, and new paradigms will take shape, all contributing to the endless cycle of knowledge entropy. In this context, the ability to effectively manage and navigate the EoN will be increasingly critical to the success and impact of scientific research. This will require a sustained commitment to the principles and practices outlined in this paper, from the cultivation of intellectual humility and the promotion of interdisciplinary collaboration to the strategic allocation of resources and the continuous updating of research methods and standards. It will also demand a willingness to welcome the inherent uncertainty and provisionality of scientific understanding, recognizing that the pursuit of knowledge is an ongoing and open-ended endeavor. By accepting the EoN as a natural and necessary part of the scientific process, we can develop a more resilient and adaptable approach to the challenges of discovery.

While the EoN framework offers a promising new perspective on the nature of scientific discovery, it is also clear that much more research and reflection are needed to fully elaborate and apply these ideas. There are many open questions and avenues for further exploration, from the development of quantitative measures of knowledge entropy to the investigation of its implications for specific scientific fields and research programs. EoN raises deeper questions about the nature of truth, the limits of human understanding, and the role of science in society. These are questions that have long been the subject of philosophical and sociological inquiry, and that take on new urgency and significance in light of the challenges posed by the EoN.

One potential criticism of the EoN framework is that it may be seen as overly pessimistic or defeatist, suggesting that the pursuit of knowledge is inherently fraught with uncertainty and error. Some may argue that this perspective could undermine the confidence and motivation of researchers, or lead to a form of relativism that questions the very possibility of scientific progress. To address these concerns, it's important to emphasize that the EoN framework is not intended to deny the reality of scientific progress or the value of the scientific enterprise. Rather, it seeks to provide a more realistic and nuanced understanding of the challenges involved in the pursuit of knowledge, and to highlight the importance of strategies and practices that can help navigate this complexity.

Another potential limitation is that entropy, as borrowed from thermodynamics and information theory, may not fully capture the specific dynamics and characteristics of knowledge production. There may be important differences between physical systems and epistemic systems that limit the applicability of entropy as a metaphor or explanatory framework. To mitigate this, it's crucial to acknowledge the limitations of the analogy and to use entropy as a heuristic tool rather than a strict scientific model. The EoN framework should be seen as a conceptual lens that can provide valuable insights and generate new hypotheses, rather than a complete or definitive theory of scientific progress.

This paper is intended not as a final word on the subject, but as an invitation to further dialogue and investigation. Only by engaging with the full richness and complexity of the world that we can hope to deepen our understanding of it, and to realize the boundless potential of the human mind. By bringing together insights from a range of disciplines and perspectives, from thermodynamics and information theory to the history and philosophy of science, we hope to stimulate new ways of thinking about the complexities of scientific discovery and the pursuit of knowledge more broadly.

Parrinder, P. (2015). *Utopian Literature and Science: from the scientific revolution to Brave New World and beyond*. Springer.